**Editor's Note:** In an investigation of student problem solving, the authors discovered that students often chose to answer problems with their gut feelings, rather than performing a calculation or supporting their answer with conceptual reasoning. This work explores the reasons for this choice, and the paper concludes with recommendations for improvements to physics instruction.

**Investigation of student and faculty problem solving: An example from quantum mechanics**

Alexandru Maries[1], Ryan Sayer[2], and Chandralekha Singh[3]

[1]Department of Physics, University of Cincinnati, Cincinnati, OH 45221, USA
[2]Department of Physics, Bemidji State University, Bemidji, MN, 56601, USA
[3]Department of Physics and Astronomy, University of Pittsburgh, Pittsburgh, PA, 15260, USA

**Abstract**

We describe a study focusing on students' and faculty members' reasoning about problems of differing cognitive complexity related to the double-slit experiment (DSE) with single particles. In the first phase of the study, students in advanced quantum mechanics courses were asked these questions in written form. Additionally, individual interviews were conducted with ten students in which they were asked follow-up questions to make their thought processes explicit on the challenging problems. Students did well on the straightforward problem, showing they had some knowledge of the DSE after traditional instruction, but they struggled on the more complex ones. Even if explicitly asked to do so in interviews, students were often uncomfortable performing calculations or making approximations and simplifications, instead preferring to stick with their gut feeling. In the second phase of the study, the problems were broken down into more pointed questions to investigate whether students had knowledge of relevant concepts, whether they would do calculations as part of their solution approach if explicitly asked, and whether they explicitly noted using their gut feeling. While the faculty members' responses suggest that they could seamlessly move between conceptual and quantitative reasoning, most students were unable to combine concepts represented by different equations to solve the problems quantitatively. We conclude with instructional implications.

**Introduction**

The double slit experiment (DSE) with single particles can improve student understanding of foundational concepts of quantum mechanics while simultaneously helping them learn to "think like a physicist". Broadly speaking, to think like a physicist, one should use physics knowledge flexibly to reason about problems [1-4]. In particular, research has often focused on how physics experts reason, e.g., in the context of solving problems to understand their problem-solving, reasoning and meta-cognitive processes and contrast them with those of students in order to provide instructional guidelines [3,4].

In mathematics, Schoenfeld emphasized the importance of students learning to think mathematically [5,6] describing it as: "you understand how to think mathematically when you are resourceful, flexible, and efficient in your ability to deal with new problems in mathematics" [5]. Price et al. conducted interviews with 52 expert scientists and engineers in which they used authentic problems in their field of expertise and developed a framework involving 29 decisions that experts use when solving such authentic problems [3]. The authors emphasize the importance of giving students practice and feedback with making these decisions while solving problems in their undergraduate curriculum to help them progress towards becoming experts.

In quantum mechanics, the nature of the concepts can place an additional burden to developing expertise [7-11]. Research shows that in introductory classical mechanics, students' everyday notions of how the world works interfere with learning physics, and in quantum mechanics, students' knowledge of concepts such as position, momentum, energy from classical mechanics interfere with learning how to reason about these concepts in quantum mechanics [11]. However, even in quantum mechanics, students can be provided opportunities to help them develop their problem solving and reasoning skills simultaneously while learning these new concepts by providing them with practice and feedback.

Here, we discuss the second phase of an investigation about how students in modern physics and advanced quantum mechanics as well as physics faculty reason about the DSE when solving problems of varying cognitive complexity. Findings from the first phase of the study were summarized in a conference proceedings [12], and we will begin by reviewing the main findings from the first phase and how they informed the design of the second phase of this investigation. We found that, instead of reasoning systematically, many students resorted exclusively to guessing without using either conceptual reasoning or calculations, which we will refer to as relying on their "gut feeling." During the second phase of the study, the problems were refined into more specific questions to examine whether students possessed knowledge of relevant concepts and whether they could perform calculations as part of their solution if explicitly prompted. We found that a majority of students were unable to perform more complicated calculations that required combining more than one concept. We contrast student reasoning on these problems with those of physics faculty members who moved flexibly and seamlessly between conceptual and quantitative reasoning to check their intuition and asked themselves good questions while navigating the problem-solving process, thus providing a concrete example of what thinking like a physicist may look like in practice. We end with recommendations for how physics educators can help their students learn to think like a physicist and develop expert-like problem-solving and reasoning skills while simultaneously helping them learn physics concepts.

**Performance of students and faculty: Phase I**

In phase I of the study, 46 students in an upper-level quantum mechanics (QM) course were administered the following problems about double-slit interference after the relevant instruction had been completed.

> Q1: You are conducting a double-slit experiment in which you send a large number of non-relativistic electrons of the same kinetic energy one at a time towards a double-slit plate. The slit width is 50 pm, the slit separation is 1 nm and the distance between the slits and the screen is 3 m. If the wavelength of the electrons is 9 pm, describe the pattern you expect to observe on the screen after a large number of electrons have passed through. Explain your reasoning.

Students performed well (78% of answers were deemed "acceptable") and in the written responses, many noted that there will be an interference pattern on the screen because the wavelength of the electrons is comparable to the slit width. The next question asked about substituting protons for electrons:

> Q2: Suppose the experiment is modified by using protons instead of electrons while all the following parameters are held fixed: kinetic energy, slit width, slit separation, distance between slits and screen. How does the pattern change, if at all? Explain your reasoning.

On this problem, 34% of the QM students answered correctly that the fringe spacing decreases. However, the majority of these students seemed to use their gut feeling because they either did not provide reasoning or their reasoning amounted to "protons have larger mass, therefore they have smaller wavelength". Follow-up interviews confirmed that students often used their gut feeling to answer the question, rather than reasoning physically.

The next problem replaced protons with grains of sand:

Q3: Consider particles of sand, which can be approximated as spheres of a radius of about 1/10 of a millimeter. Do you expect that a double slit experiment with well-chosen parameters would show an interference pattern? Explain your reasoning.

On this problem, the QM students rarely attempted estimating the wavelength of sand particles to help them reason through this problem (only one out of the 46 students providing written responses (2%) did so). However, on the same quiz, nearly all students answered correctly when asked a straightforward question on calculating the de Broglie wavelength.

Following this written study, individual interviews were conducted with six QM students and five faculty members using a semi-structured, think-aloud protocol to get a better understanding of their reasoning. They were not explicitly asked to explain their reasoning by using calculations. Four additional QM students were interviewed and only asked Q3 to understand their reasoning.

In contrast to the students, the responses of interviewed faculty members indicate that they reasoned quite deliberately, and even though they made use of intuition, they also double checked it by making approximations and simplifications and using them to carry out calculations (or estimations). For example, when asked about the interference pattern formed by protons, a faculty member drew on knowledge of a number of relevant equations ($K = p^2/2m$, $\lambda = h/p$, $d\sin\theta = n\lambda$) and fluently inserted numbers and performed rough calculations.

When asked about the interference of sand particles, faculty members generally noted that sand particles are so large that their wavelength would be unobservable, but they could also explicitly show via approximations, estimations and calculations that designing a DSE experiment in which the sand particles would show interference is not possible. For example, one faculty member, while not performing any calculations, stated that the wavelength would be so small that the pattern would shrink to become unobservable.

Equally importantly, the interviews with faculty members suggest that they knew what and where the relevant knowledge was in their knowledge structure so that they could create a more refined solution if needed, e.g., by using the de Broglie relation. In addition, their responses often included making estimations not just of the wavelength, but also of the kind of DSE that one might have to create to be able to observe interference to show that it is not possible with sand particles. For example, below we provide an excerpt from an interview with a faculty member (I = interviewer, FM = faculty member):

> FM: "I don't think I would [expect to be able to design a DSE experiment to observe interference]. 1/10[th] of mm-that is 100 micrometers. This is of the order of $10^{20}$ protons. That is how big you expect the sand particle to be… The mass of the sand particle is about $10^{20}$ larger than proton… The short answer is no, realistically no… it would have to be the wavelength lambda divided by slit width times the separation to the screen… If the fringes show up that would be roughly the order of the distance you will expect. First off all this number is going to be ridiculously small… Suppose you manage to achieve all the criteria for quantum coherence in a double slit experiment which means that it is cold enough that you manage to keep quantum coherence, forget about the internal degrees of freedom, hypothetically we will be able to do all of that, even then your distance between fringes will be so small that optically the fringes won't be resolvable…Oh, I forgot, I guess in double slit, the distance between the slits will be dominant one but on top of that there is single slit pattern.…The other thing is that normally you wouldn't expect it to be quantum coherent because each sand particle consists of so many internal particles that you also need to make it really really cold so that you treat it as one quantum particle as opposed to a many body system to actually get interference so everything will be $10^{20}$ bigger and $10^{20}$ harder […] to do this experiment because the de Broglie wavelength will be $10^{20}$ smaller than in the previous case."

> I: "What will happen if you let the particles go through the slit?"

> FM: "It won't even fit through the slit, so you will need quantum tunneling on top of that in order to make it work."

I: "What will happen if the slit width is of the size of the particles?"

FM: "Well, in an actual double slit or single slit experiment… that would make the interference pattern even smaller… You will run into more practical problems. Having a big slit means that you will have more practical problems…"

I: Can you have the de Broglie wavelength be smaller than the size of the object?

FM: "Yes…well it does not mean you will not see interference pattern. It just means it is going to be that much harder to see it. Like there is no sharp cutoff that if the de Broglie wavelength is smaller than the separation of atoms in the sand then you will not see interference pattern. But the short answer is that you should treat it more classically than quantum. I imagine the visibility will be heavily suppressed to the point that it would be impossible to see interference… I am saying you should not be able to see patterns because the numbers are such that it is not practical to see double slit pattern. It does not mean quantum mechanics does not apply to particles of this size. It is just that you do not expect to observe any quantum interference at all–you would expect to see the same result as if you had done the calculation classically which means in your case it will just go through one of the slits… it will be unresolvable."

This interview with the faculty member suggests that they were thinking like a physicist and resorting to metacognition to think through the situation on their feet when probed further. They was able to resort to their knowledge of physics as appropriate to refine his arguments while problem-solving.

In the appendix, we provide another example from an interview with a faculty member, and we also make an explicit comparison with the Price et al. study to identify which specific problem-solving decisions describe by Price et al. were used by the faculty members in reasoning through this question [3] .

**Performance of students: Phase II**

Since in phase I we found that students had difficulty with Q2, we wanted to understand whether this was because they did not have the prerequisite knowledge, or perhaps they had the prerequisite knowledge but did not apply it, or they had the prerequisite knowledge, they tried to apply it, but did so incorrectly. Therefore, in phase II of the study, we first asked students questions about each piece of prerequisite knowledge. These students were enrolled either in a modern physics (MP, 48 students) or an upper-level undergraduate quantum mechanics (QM, 28 students) course, and they answered the questions in class during the last week of classes, after traditional lecture-based instruction in relevant concepts. Students were provided extra credit (problems were graded for completeness).

The questions on the pre-requisite knowledge were:

**Qa**. Write down the de Broglie relation that relates the momentum of a particle ($p$) with its wavelength ($\lambda$).

**Qb**. Consider a double-slit experiment in which you send single electrons of wavelength $\lambda$ through a double slit. The distance between the slits is $d$ and the angle that the $n^{\text{th}}$ bright fringe makes with the midpoint between the slits is $\theta$. Write the relation between $d$, $\lambda$, $n$, and $\theta$.

**Qc.** What is the relation between the kinetic energy ($K$), magnitude of momentum ($p$), and mass ($m$) of a particle?

We found that roughly 48% of MP students and 61% of QM students had all the prerequisite knowledge (see Table I). Assuming that students from different years had similar knowledge, we conclude that some students in the phase I study were unable to perform calculations because they did not have all the prerequisite knowledge.

Table 1. Percentages of students from modern physics (MP) and quantum mechanics (QM) who answered Qa, Qb, and Qc correctly, as well as those who answered all three questions correctly (last row).

|  | MP (N=48) | QM (N=28) |
|---|---|---|
| Qa (de Broglie) | 70% | 82% |
| Qb ($d \sin \theta = n\lambda$) | 77% | 79% |
| Qc ($K = p^2/2m$) | 79% | 100% |
| Qa, Qb & Qc | 48% | 61% |

Following these preliminary questions, we asked students Q2 from phase I, except that a final sentence was added: "Explain your reasoning using equations." We found that, similar to phase I, students struggled to answer this question correctly: 28% for MP and 33% for QM (a nearly identical percentage to phase I for QM students). The details of how many students chose each answer category are given in the supplementary material.

With regard to *why* students struggled on this question, we find the following in MP: Only roughly half the students had the prerequisite knowledge (23 out of 48), and among the students who did not have the prerequisite knowledge, the majority (18 out of 25, or 72%) answered the question incorrectly. Out of the 23 students who had the prerequisite knowledge, only 9 students (39%) attempted to use all three equations, and 1 student (4%) used only some of the three equations. Another 9 students (39%) appeared to use their gut feeling as their answers did not refer to equations (despite being explicitly asked to explain their reasoning using equations). For 3 students, we are unable to tell if they can use the equations correctly because they either said there would be no change (in one case because the student thought the masses are the same, and in the other, the student said that the equations don't depend on charge so there wouldn't be a change) or they said there would be a change but did not specify what kind of change. Out of the 9 students who used all three equations, 5 students used them correctly, 3 made a mistake, and for one, we cannot tell because the student noted there would be a change but did not specify what kind of change. Lastly, one student did not answer Q2.

In QM, 17 students (61%) had all the prerequisite knowledge. For the ones who did not have the prerequisite knowledge (11 students), the majority answered the question incorrectly (9 students). Out of the 17 students who had the prerequisite knowledge, only 8 (47%) attempted to use all three equations, and 5 students (29%) only used some of the three equations. Another 3 students (18%) appeared to use their gut feeling and one student did not answer the question. Out of the 8 students who attempted to use all three equations, 3 used them correctly and 5 made a mistake. So even among the QM students who had all the prerequisite knowledge *and* attempted to use it, most were unable to use it correctly (though, of course, the numbers are small).

In the supplementary material, we provide sample answers from QM students in each category. Here we quote students in the category that is most interesting to us: those who had all the prerequisite knowledge but did not use it:

- "The proton does not have this 'wave particle' duality so [it won't interfere]" (We note that the student drew a picture which indicated that the protons would gather in two specific spots rather than show interference.)
- "Proton also has a wavelength (de Broglie) but since its more massive, its wavelength is small so the pattern will be spread."

In short, MP and QM students struggled with this question because 1) many lacked the prerequisite knowledge (based upon their response to Qa-Qc), 2) if they had the prerequisite knowledge, they often did not use it, and 3) even if they had the prerequisite knowledge *and* used it, they often made mistakes. Out of the 23 MP students who had the prerequisite knowledge, 5 students (22%) gave correct solutions supported by calculations, and out of the 17 QM students with the prerequisite knowledge, 3 students (18%) gave correct solutions supported by calculations.

We also explored why students in phase 1 struggled to show whether it is possible to observe double-slit interference of sand grains. In phase II of the study, we divided this question into two parts:

> **Q3a.** Consider particles of sand, which can be approximated as spheres of a radius of about 0.1 millimeters. Estimate the wavelength of a sand particle. You can approximate any other parameters you need.

In MP, 33 students (69%) knew the de Broglie relation. Out of these students, 16 (48%) tried to do a calculation to estimate the wavelength of a particle of sand, and 5 (15%) tried to do a calculation, but they did not use the de Broglie relation. The other 12 students (36%) did not attempt to do a calculation and just said that the wavelength is either small or large. Out of the 16 students who tried to do a calculation, 11 students (69%) got a final numerical answer. These answers are quite varied: one is 500 m, six are between $10^{-24}$ m and $10^{-29}$ m, and the other four are even smaller with the lowest being of the order $10^{-41}$ m. A 0.1 mm sand grain has mass of about $10^{-8}$ kg, so traveling at 1 m/s, its wavelength is $\sim 10^{-25}$ m, so 6 out of 33 students (21%) who knew the de Broglie relation were able to calculate a reasonable answer. The most common difficulties were struggling to estimate the mass of a grain of sand and not knowing Planck's constant.

The QM students faired arguably worse: 23 students (82%) knew the de Broglie relation, but only 4 of these students (17%) attempted to do a calculation using the de Broglie relation (and three of the four used sound reasoning, though their answers were off due to them not knowing the correct value for Planck's constant; the fourth student's reasoning is hard to follow and their answer is significantly off). One student tried to do a calculation, but did not use the de Broglie relation. Among the rest, 13 students (56%) did not attempt to do a calculation at all and just said that the wavelength is either small or large, and 5 students (21%) did not answer the question.

Comparing the two groups, students in MP were more likely to attempt to use the de Broglie relation to estimate the wavelength of a grain of sand compared to QM students (48% vs. 17%), and conversely, MP students were less likely than QM students to ignore their knowledge of the de Broglie relation and use their gut feeling (36% vs. 56%). It is difficult to ascertain why this is the case, but one hypothesis is that students in QM are less likely to do calculations that lead to numerical answers in a QM class compared to students in MP. So QM students may be less attuned to attempting numerical calculations compared to MP students.

In the supplementary material, we provide some examples from QM students who did not get a final numerical answer for the wavelength of a particle of sand or did not attempt to perform a calculation.

Next, students answered Q3b:

> **Q3b.** For Q3a, do you expect that a double slit experiment with well-chosen parameters would show an interference pattern? Use an explicit calculation to prove that such an experiment is possible or not possible.

On this question, 65% of the MP students and 76% of the QM students noted they don't expect that a double-slit experiment can show interference for particles of sand. However, very few of these students attempted to do a calculation to explain their reasoning: 15% for MP (6 students out of the 40 who answered this question) and 10% for QM (2 out of 21). What is interesting is that in MP, we saw that 11 students calculated a wavelength and nearly all of them got an answer of the order $10^{-24}$ m or smaller. But only 5 of these students attempted to do some sort of calculation to prove that interference is not possible, despite being explicitly asked to do so, so we see that even students who had already done most of the difficult work (estimating the wavelength of a grain of sand) did not perform calculations to support their gut feeling.

This conclusion is bolstered by students' answers to Q4 (shown in Table 2) where they were asked to indicate how they had approached Q3b. Table 2 shows that the vast majority of students in both MP (81%) and QM (86%) used their gut feeling when answering Q3b because they did not know how to perform a calculation. It is also remarkable how similar the percentages are between the two groups! We also note that among students who selected C in MP, most did not actually do a calculation to show their work (out of the eight students who selected C in Q4, only two did a calculation in Q3a.) They might, for example, have chosen C not because they knew how to perform the calculation, but because they did not have a gut feeling.

Table 2: Percent of students in modern physics (MP) and quantum mechanics (QM) who selected each answer choice on Q4. The *N*'s indicate the number of students who answered the question.

|  | MP (N = 43) | QM (N = 22) |
|---|---|---|
| A. I have a gut feeling that it is not possible to see interference pattern but I do not know how to prove it because I don't know what calculations to perform. | 49% | 55% |
| B. I have a gut feeling that it is possible to see interference pattern but I do not know how to prove it because I don't know what calculations to perform. | 33% | 32% |
| C. I don't agree with either A or B. | 19% | 14% |

**Discussion and Summary**

This study started with the question of what an expert solution to a challenging physics problem would look like (faculty solutions) and how student solutions would compare to these expert solutions. We observed in phase 1 that very few students were able to approximate these expert solutions; for example, only one out of 46 students attempted to calculate the de Broglie wavelength of a sand grain. Many simply used a gut feeling to answer the questions instead of performing a calculation. Q4 explained this reliance on gut feelings: at least 81% of MP students and 86% of QM students did not know what calculation to perform. Questions Qabc gave some explanation for this inability: only 48% of MP students and 61% of QM students knew the basic relationships that would enable them to perform a calculation. Additional explanations came from the phase 2 version of Q2 that explicitly asked for a calculation: 1) even for those students who answered Qabc correctly, and thus were not only aware of these relationships but also were freshly reminded of them, only 39% (MP) and 47% (QM) of students attempted to use all three relationships, and 2) students who attempted to use all three relationships frequently made mistakes. For those students who knew all the correct equations and yet did not attempt a calculation, we are unable to conclude whether they really didn't know how to combine them to find an answer or whether they simply preferred to stick with their gut feeling without testing it.

**Implications for Instruction**

Recognizing the types of reasoning used by faculty members to solve complex problems is the first step in designing instruction to help students develop expertise and learn how to apply some of the decisions needed to solve authentic problems [3]. In the context of the DSE problems discussed in this study, when focusing on faculty members' responses, there are two important findings:

1. While faculty members generally appeared to have reasonably good intuition about problem solutions, they made use of multiple problem-solving decisions to check their intuition and showed flexibility in moving between quantitative and qualitative reasoning.
2. In the less-structured problem about whether sand particles will show a DSE interference pattern with suitable parameters, faculty members engaged in meta-scaffolding, i.e., they were able to scaffold their own thinking and reasoning by asking themselves the right kinds of questions and what they may need to approximate to arrive at a conclusion about a physical situation (e.g., that would help them get clarity on problem solution, i.e., no interference pattern can be observed for particles of sand in any realistic situation).

Based upon student performance on DSE problems, the expertise of students in this study spanned a spectrum with most unable to complete the multi-part reasoning required to answer Q2. We note that the majority of students who knew the basic relationships required to reason on Q2 did not use them and instead stuck with their gut feeling. If students stick with using only their gut feeling instead of using equations to check their gut feeling and if they lack the ability to use the equations in multi-step reasoning, it is not surprising that they were not able to answer Q3, since its absence of structure (no specified slit width/distance, screen distance, etc.) made the multi-part reasoning even more challenging, thus further encouraging students to stick with their gut feeling.

In the $21^{st}$ century, when the ability to change and learn is paramount in careers [13], problem-solving and reasoning skills will be as important as knowing any specific content. Thus, students must be given grade incentives and opportunities for solving multi-part problems as well as less structured and out of the box problems in which they must make approximations, decide on relevant sub-problems, and carry out calculations (or estimations) to show why something is possible or not instead of the assessment methods emphasizing solving problems algorithmically. Physics content should be used as a tool for developing student expertise and helping them learn to make the kinds of problem-solving decisions that are routinely carried out by experts [3]. Montgomery et al. found [4] that a research-based capstone course which was designed to give students an experience that would simulate an authentic research environment gave students the opportunity to make many more expert-like decisions. The authors also noted that it may be possible for courses to provide more practice with these problem-solving decisions if they "involve a research element or include real-world problems that don't provide all necessary information up front" [4]. Similarly, using projects in physics courses could be beneficial, particularly if they require group work with dedicated roles (more below) and focus on students using the skills and content knowledge gained in the course to investigate a scientific question or phenomenon.

In an introductory physics context, some have pointed out that traditional problems that physics instructors typically give to students in homework, quizzes and exams do a poor job of helping students with the kinds of reasoning needed to solve authentic problems and learn to think like a physicist [3,4,14]. However, even in advanced courses, traditional physics problems are very often algorithmic and perfectly specified, i.e., they provide exactly the information needed to solve the problem, nothing more, nothing less, in addition to being specific about what students need to calculate [4]. In other words, they do not provide students with opportunities to make decisions about what information is relevant and what is extraneous as well as about what specifically to calculate. While solving algorithmic physics problems can help students learn basic procedures, if they are the only types of problems the course assessment focuses on, students are unlikely to learn how to think like a physicist and develop the skills necessary to solve authentic problems that are going to be common in their future careers as scientists [3].

Consequently, to help students learn to think like a physicist, it is important for physics instructors to help students engage with problems of different cognitive complexity with some being less structured than others [15,16]. Previously, in the context of introductory physics, some have advocated for using "context-rich problems" [17], which are often ill-specified problems related to real-world situations, where "ill-specified" means that the problems require students to make decisions about what to calculate, sometimes also requiring students to make reasonable estimates of things that are not known. Q3 is an example of such a problem. In addition, the approach they advocate for (which can be beneficial with tasks described here as well) includes using different roles which rotate from week to week – Manager, Skeptic, Checker/Recorder which "reflect the mental planning and monitoring strategies that individuals must perform when solving problems alone." [18] In other words, using the different roles can help students develop expertise with making some of the decisions outlined earlier in the article.

Since helping students learn to solve problems should be a major goal of physics curricula at all levels, it is important to make it part of the learning objectives and course assessment. Physics educators can learn from Schoenfeld [6] who noted, "to develop the appropriate mathematical habits and dispositions of interpretation and sense-making as well as the appropriately mathematical modes of thought… the communities of practice in which they learn mathematics must reflect and support those ways of thinking. That is, classrooms must be communities in which mathematical sense-making, of the kind we hope to have students develop, is practiced." Schoenfeld demonstrated that students developed appropriate mathematical habits and disposition when he structured his math courses such that students worked in small groups on challenging problems and he would go to each group and ask them questions such as, what are you doing, why are you doing it, how does it get you closer to your goals, etc. In addition, to show students what authentic problem solving looks like and that struggling is a normal part of problem solving, students in his courses had the option to pose challenging problems to him "cold" in order to see how he dealt with them. Often he was familiar with the problems, and he told them so. But occasionally, an unfamiliar problem proved truly challenging. After working on it unsuccessfully for 5-10 minutes, he stopped and asked his students to keep thinking about the problem and let him know if they found a solution before the next class meeting. When they next met, they discussed his solution and theirs. Physics educators can adopt similar strategies to help their students learn to apply appropriate problem-solving decisions.

While having students engage in small groups with challenging physics problems, students may become overwhelmed, i.e., a multi-part problem can lead to cognitive overload due to students' limited information processing capacity [19]. Physics instructors should also recognize that the dichotomy between helping students learn physics concepts and developing their problem-solving skills is false and they must do both of these simultaneously throughout their instruction. Therefore, it is important to provide appropriate scaffolding support to help students break a multi-part problem into sub-problems [20]. At first, students can be given multi-part problems where the separate parts are scaffolding support intended to show them how to break up a more complex problem into sub-problems, as well as to help reduce their cognitive load while problem-solving so they are able to develop some of the skills necessary for solving multi-part problems. After they show sufficient facility with problems that involve this kind of scaffolding support, the support should be gradually reduced to help them develop self-reliance [20] so that at the end, they are able to decide on how to break up a multi-part problem on their own without any guidance. Furthermore, for such approaches to be effective, they should be incorporated in an instructional context which uses multiple pedagogical considerations that are consistent with supporting student learning more generally, e.g., see Tables II and III in Ref. [14]. We also note that technology can be very useful in creating a course structure which ensures that in-class time is used for giving students practice with solving problems that require making many of the 29 decisions to help them develop expertise. For example, some of the lower-level content (e.g., definitions of concepts, simple relationships) can be moved to outside of class via pre-lectures such that more in-class time is available for problem solving in which students learn to apply concepts in diverse situations. Finally, developing facility with making these expert-like decisions in our students is a continuous goal. Developing expertise is a process and students need to strengthen their problem-solving and reasoning skills by being given opportunities and grade incentives to solve challenging problems that require expertise beyond algorithmic facility with appropriate scaffolding support.

**Acknowledgements**

We thank the NSF for award PHY-2309260. We also thank all the students and faculty who participated.

**Author declarations**

Conflict of interest: The authors do not have any conflicts of interest to disclose.

**References**

[1] C. Singh, When physical intuition fails, American Journal of Physics **70**, 1103 (2002).
[2] A. Van Heuvelen, Learning to think like a physicist: A review of research-based instructional strategies, American Journal of Physics **59**, 891 (1991).
[3] A. M. Price, C. J. Kim, E. W. Burkholder, A. V. Fritz, and C. E. Wieman, A Detailed Characterization of the Expert Problem-Solving Process in Science and Engineering: Guidance for Teaching and Assessment, CBE—Life Sciences Education **20**, ar43 (2021).
[4] B. J. Montgomery, A. M. Price, and C. E. Wieman, How traditional physics coursework limits problem-solving opportunities, Phys. Educ. Res. Conf. Proc., 230 (2023).
[5] A. H. Schoenfeld, Theoretical and pragmatic issues in the design of mathematical problem solving instruction, Paper presented at the annual meeting of the American Educational Research Association (Montreal, Quebec, Canada, Apr. 11-14, 1983)  (1983).
[6] A. H. Schoenfeld, Learning to Think Mathematically: Problem Solving, Metacognition, and Sense Making in Mathematics (Reprint), Journal of Education **196**, 1 (2016).
[7] C. Singh, Student understanding of quantum mechanics, Am. J. Phys. **69**, 885 (2001).
[8] M. C. Wittmann, R. N. Steinberg, and E. F. Redish, Investigating student understanding of quantum physics: Spontaneous models of conductivity, American Journal of Physics **70**, 218 (2002).
[9] C. Singh, Interactive learning tutorials on quantum mechanics, American Journal of Physics **76**, 400 (2008).
[10] A. Kohnle, I. Bozhinova, D. Browne, M. Everitt, A. Fomins, P. Kok, G. Kulaitis, M. Prokopas, D. Raine, and E. Swinbank, A new introductory quantum mechanics curriculum, European Journal of physics **35**, 015001 (2013).
[11] E. Marshman and C. Singh, Framework for understanding the patterns of student difficulties in quantum mechanics, Physical Review Special Topics-Physics Education Research **11**, 020119 (2015).
[12] R. Sayer, A. Maries, and C. Singh, Advanced students' and faculty members' reasoning about the double slit experiment with single particles, in *Proceedings of the Physics Education Research Conference* (2020).
[13] See supplementary material online for another example of a faculty member thinking like a physicist, a comparison of the faculty members' reasoning with the Price et al. framework, a detailed breakdown of students' answers to Q2 from phase 2, and sample student answers to Q2 and Q3.
[14] L. McNeil and P. Heron, Preparing physics students for 21st-century careers, Physics Today **70**, 38 (2017).
[15] M. Dancy and C. Henderson, Framework for articulating instructional practices and conceptions, Phys. Rev. ST Phys. Educ. Res. **3**, 010103 (2007).
[16] M. Good, E. Marshman, E. Yerushalmi, and C. Singh, The value of using different types of physics problems to help students become proficient problem-solvers, Phys. Educ. **59**, 015018 (2023).
[17] C. Wieman, Comparative Cognitive Task Analyses of Experimental Science and Instructional Laboratory Courses, Phys. Teach. **53**, 349 (2015).
[18] P. Heller and M. Hollabaugh, Teaching problem solving through cooperative grouping. Part 2: Designing problems and structuring groups, American Journal of Physics **60**, 637 (1992).
[19] P. Heller, R. Keith, and S. Anderson, Teaching problem solving through cooperative grouping. Part 1: Group versus individual problem solving, American journal of physics **60**, 627 (1992).
[20] F. Paas, A. Renkl, and J. Sweller, Cognitive Load Theory: Instructional Implications of the Interaction between Information Structures and Cognitive Architecture, Instructional Science **32**, 1 (2004).

[20]    A. Collins, J. S. Brown, and S. E. Newman, Cognitive Apprenticeship: Teaching the Crafts of Reading, Writing, and Mathematics, in *Knowing, learning, and instruction: Essays in honor of Robert Glaser.* (Lawrence Erlbaum Associates, Inc, Hillsdale, NJ, US, 1989), p. 453.

# Supplementary material

*Another example of thinking like a physicist from an interviewed faculty member:*

The faculty member started by stating that a sand particle is macroscopic and essentially behaves classically due to decoherence. However, they stated that assuming we ignore the decoherence effects and pretend that we can treat the sand particles quantum mechanically:

> "then interference effects will depend on the de Broglie wavelength which will be incredibly small for a grain of sand. So small that you will never be able to construct a grating. The minimum size of the grating will be the size of the slit so you can calculate basically where the first order diffraction peak will be and what the angle will be. And my guess is that you would not resolve it within the size of the universe. [...] Let me estimate the de Broglie wavelength for a sand particle moving at a speed of 100 m/s. If the mass of the sand particle is $10^{-5}$ kg then its momentum is p = mv = $10^{-3}$ kg m/s. So the de Broglie wavelength is $\lambda = h/p = 6 \times 10^{-34}$ J s/$10^{-3}$ kg m/s = $10^{-30}$ m."

The faculty member then drew a diagram showing a point on the distant screen and connected it to the center of the two slits and drew an angle theta for the angle between them and stated that:

> "in order to see first order fringe on a very distant screen, $d \sin\theta = \lambda$ or $d\theta = \lambda$ where let's assume that the distance between the slits d is of the order of the size of the sand particle $10^{-4}$ m. Note that if we make d larger than this, the angle theta will be even smaller and resolving two bright fringe on the screen will become even more difficult... So the angle $\theta = \lambda/d = 10^{-26}$. The question is... how far should the screen be where interference fringes are observed so that two adjacent bright fringes, say zeroth order and first order can be resolved? The distance between adjacent fringes should be at the very least of the order of the size of the sand particles $10^{-4}$ m so the minimum distance of the screen from the two slits in order to resolve it should be $L = 10^{-4}$ m/$\theta$ =$10^{-4}$ m/ $10^{-26} = 10^{22}$ m."

The faculty member then estimated the size of the universe by using its age and the speed of light to obtain ~$10^{26}$ m. He then went on to say:

> "$10^{-4}$ times size of the universe, -it is a very very big distance. I am not an astrophysicist, but people work in astronomical unit which is the distance between earth and sun... so 1/10000 times the size of the universe [...] is a very very large distance that is not practical for doing a double slit experiment with distant screen. The problem is that we have a universe filled with objects and radiation and those are going to distinguish between different paths because the universe is not empty. There is no way you will be able to do this experiment with parameters of the sand and see interference patterns".

Again, the interview displays how the faculty member displayed an important aspect of what it means to think like a physicist. Their metacognitive process focused on showing that it is not possible to resolve the fringes within the size of the universe.

*Comparison with the Price et al. framework*

Comparing with Price et al. [1], it is instructive to identify which of the 29 decisions faculty made while engaged in problem solving to answer Q3. In Table 1, we list all the different decisions made by faculty along with examples from the data. We note that not all the decisions were made by each of the faculty members, but regardless, they used a lot more than the students who primarily relied on a gut feeling and were not able to do a calculation even when explicitly asked. This is in stark contrast to the students who primarily made a prediction using their gut feeling but typically did not make most of the problem-solving decisions in the list above to check their prediction (81% in MP and 86% in QM).

**Table 1.** Examples from faculty interviews of the problem-solving decisions [1] made by faculty when solving Q3.

| Problem solving decisions | Examples from faculty interviews |
|---|---|
| Choosing the goals, criteria, and constraints | "The question is… how far should the screen be where interference fringes are observed so that two adjacent bright fringes, say zeroth order and first order can be resolved?" |
| Considering the important underlying features or concepts that apply | "The distance between adjacent fringes should be at the very least of the order of the size of the sand particles $10^{-4}$ m so the minimum distance of the screen from the two slits in order to resolve it should be $L = 10^{-4}$ m/$\theta$ = $10^{-4}$ m/$10^{-26}$ = $10^{22}$ m." |
| Considering how to narrow down the problem | "Suppose you manage to achieve all the criteria for quantum coherence in a double slit experiment which means that it is cold enough that you manage to keep quantum coherence, forget about the internal degrees of freedom, hypothetically we will be able to do all of that, even then your distance between fringes will be so small that optically the fringes won't be resolvable" |
| Considering which predictive framework to use | "it would have to be the wavelength lambda divided by slit width times the separation to the screen… If the fringes show up that would be roughly the order of the distance you will expect" |
| Considering how to decompose the problem into sub-problems [even if done implicitly without the planning being made explicit] | "interference effects will depend on the de Broglie wavelength" […] "Let me estimate the de Broglie wavelength" […] "So the de Broglie wavelength is … $10^{-30}$ m" […] "in order to see first order fringe on a very distant screen, $d \sin \theta = \lambda$ or $d\theta = \lambda$" […] "So the angle $\theta = \lambda/d = 10^{-26}$" […] "The question is… how far should the screen be where interference fringes are observed so that two adjacent bright fringes, say zeroth order and first order can be resolved?" […] "the minimum distance of the screen from the two slits in order to resolve it should be … $10^{22}$ m." |
| Considering what approximations and simplifications are appropriate | "Let me estimate the de Broglie wavelength for a sand particle moving at a speed of 100 m/s" |
| Considering what is the best way to represent and organize the information | For example, last faculty member drew a diagram to represent and organize the information. |
| Arriving at a conclusion based on the calculation | "If the mass of the sand particle is $10^{-5}$ kg then its momentum is p = mv = $10^{-3}$ kg m/s. So the de Broglie wavelength is $\lambda = h/p = 6 \times 10^{-34}$ J s/$10^{-3}$ kg m/s = $10^{-30}$ m." |
| Comparing the conclusion to the prediction | "And my guess is that you would not resolve it within the size of the universe." […] "The distance between adjacent fringes should be at the very least of the order of the size of the sand particles $10^{-4}$ m so the minimum distance of the screen from the two slits in order to resolve it should be $L = 10^{-4}$ m/$\theta$ =$10^{-4}$ m/$10^{-26}$ = $10^{22}$ m." […] "$10^{-4}$ times size of the universe, -it is a very very big distance. I am not an astrophysicist, but people work in astronomical unit which is the distance between earth and sun… so 1/10000 times the size of the universe […] is a very very large distance that is not practical for doing a double slit experiment with distant screen." |

*Detailed breakdown of students' answers to Q2 (phase 2)*

Student answers fell into six categories (see Fig. 1): fringe spacing decreases (Dec), fringe spacing increases (Inc), no change (No chg), there will be a change, but students did not specify what kind of change (Some chg), no interference (No int), and other (e.g., dark fringes become bright fringes, different interference pattern if the wavelength is different, otherwise no difference, or unclear answer). The most common incorrect answers from both MP and QM students were that the fringe spacing would increase (24% MP and 37% QM) or that there is no change (21% MP and 11% QM).

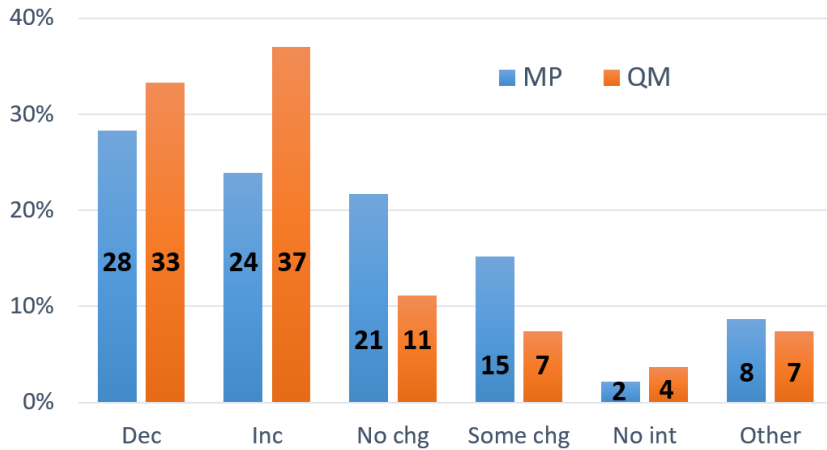

**Figure 1**. Categorization of student answers to Q2 (phase 2) and percentages of students in MP (blue – 48 total students) and QM (orange – 28 total students) in each category (decrease is correct). The numbers on the bars are the percentages of students who selected each answer choice.

*Sample answers from QM students on Q2 separated by category:*
1. Students who had all the prerequisite knowledge, but used gut feeling instead:
    - "The proton does not have this 'wave particle' duality so [it won't interfere]" (We note that the student drew a picture which indicated that the protons would gather in two specific spots rather than show interference.)
    - "Proton also has a wavelength (de Broglie) but since its more massive, its wavelength is small so the pattern will be spread."
2. Students who attempted to use all three equations but made a mistake:
    - "$m \uparrow \Rightarrow p \downarrow$ to hold $K \Rightarrow \lambda \uparrow$. Since $d \sin\theta = n\lambda$, and $\lambda$ increases, the pattern spreads out, more distance between bright spots."
        - Mistake is inferring that if mass increases, momentum decreases to keep kinetic energy the same.
    - "$K = \frac{p^2}{2m} \rightarrow p = \sqrt{\frac{K}{2m}}$. $\lambda = \frac{h}{p} = \frac{h}{\sqrt{\frac{K}{2m}}}$. Since $\lambda \propto \sqrt{m}$, the distance $d$ would have to increase as $d \propto \lambda$."
        - Mistake in solving for momentum from the equation for kinetic energy. Also, the final answer relates to the distance $d$ between the slits, but this does not change. The student appears to be confused between the distance between the slits ($d$) and the distance between adjacent maxima on the screen (which is related to the angle $\theta$).
    - "Mass is changed, increased. $K = \frac{p^2}{2m} \leftarrow p$ decreases to keep $K$ constant

$\lambda = \frac{h}{p}$ ← $\lambda$ increases since $h$ is constant

$\lambda = d \sin \theta$ ← $\theta$ has to increase.

The fringe separation will have larger spaces between them."

- o Mistake is inferring that if mass increases, momentum decreases to keep kinetic energy the same.
- "Given that the kinetic energy is constant, due to $\sqrt{2mK} = p$, the momentum will be larger by $\sqrt{m_p - m_e}$, making $\lambda$ larger by $\frac{1}{\sqrt{m_p - m_e}}$, and since $d$ = const, $\theta$ will be larger for all $n$ in $\theta = \sin^{-1} \frac{n\lambda}{d}$.
  - o Multiple mistakes, but the primary one is deriving from $\sqrt{2mK} = p$ that the momentum of the proton will be larger by $\sqrt{m_p - m_e}$. It should be larger by $\sqrt{m_p/m_e}$.

3. Students who only used a subset of the equations:
   - "The pattern doesn't change because $d \sin \theta = n\lambda$ is not dependent on mass."
   - "The wavelength of the protons is longer, so the bright fringes will be further apart because $\theta = \sin^{-1} \left(\frac{n\lambda}{d}\right)$ and $\theta$ will be greater."
   - "Since a proton is more massive than an electron, and kinetic energy is constant, $K = \frac{p_e^2}{2m_e} = \frac{p_p^2}{2m_p} \Rightarrow \sqrt{\frac{m_p}{m_e}} p_e = p_p$ and $\frac{m_p}{m_e} > 1$ so momentum of the proton is larger and hence changes the wavelength, causing a different kind of interference."

Examples on Q3a of student responses from students who did not get a final numerical answer or did not attempt to perform a calculation despite having written the de Broglie relation:
- "$\lambda = \frac{\hbar}{mv} = \frac{\hbar}{............uhhhhh}$,"
- "I cannot estimate. However, the wavelength will be <u>super large</u>" (Student emphasis)